\documentclass[aps,prb,twocolumn,groupedaddress,showpacs]{revtex4}
\usepackage{graphicx,graphics}
\begin{document}

\title{Giant thermopower and figure of merit in single-molecule devices}

\author{C. M. Finch}\email{c.lambert@lancaster.ac.uk}
\author{V. M. Garc\'{\i}a-Su\'arez}
\author{C. J. Lambert}

\affiliation{Department of Physics, Lancaster University,
Lancaster, LA1 4YB, UK}

\date{\today}

\begin{abstract}
We present a study of the thermopower $S$ and the dimensionless
figure of merit $ZT$ in molecules sandwiched between gold
electrodes. We show that for molecules with side groups, the shape
of the transmission coefficient can be dramatically modified by
Fano resonances near the Fermi energy, which can be tuned to
produce huge increases in $S$ and $ZT$. This shows that molecules
exhibiting Fano resonances have a high efficiency of
thermoelectric cooling which is not present for conventional
un-gated molecules with only delocalized states along their
backbone.
\end{abstract}


\maketitle

Recent advances in single-molecule electronics have been
underpinned by considerable improvements in electrical contacting
techniques \cite{weber2002,xu2003,Haiss2004}, which have allowed
the identification of a range of fundamental properties including
switching \cite{choi2006,lortscher2006}, rectification
\cite{ashwell2005,ashwell2006}, memory \cite{he2006} and sensing
\cite{long2006}. Single-molecule devices offer a potential route
to sub-10 nm electronics. For the purpose of designing large-scale
integration of such devices, a knowledge of thermal as well as
electrical properties is needed. Furthermore, a knowledge of
thermoelectric properties of single molecules\cite{Koh04} may
underpin novel thermal devices such as molecular-scale Peltier
coolers and provide new insight into mechanisms for
molecular-scale transport. For example from the sign of the
thermopower it is possible to deduce the conduction mechanism or
relative position of the Fermi energy \cite{paulsson2003}, with a
positive sign indicating $p$-type conduction, which means the
Fermi energy is lying closer to the HOMO level. The thermopower
can also show very interesting behaviors, like a linear increase
with the molecular length \cite{pauly2008}. Other theoretical
studies have shown the possibility of changing both the sign and
magnitude of the thermopower by electrically gating a molecular
wire \cite{wang2005,zheng2004}, which moves the Fermi energy
across a transmission resonance and thus changes its sign.

However when the Fermi energy lies within the HOMO-LUMO gap,
thermopower is expected to be low. Indeed recently, Reddy {\em et
al.} \cite{reddy2007} measured the room temperature thermopower of
1,4-biphenyldithiol, using a modified scanning tunnel microscope.
They found the thermopower to be $+12.9$ $\mu$VK$^{-1}$. This low
value was attributed to the Fermi energy sitting within the
molecule's HOMO-LUMO gap where both the DOS and transmission
coefficients are relatively flat.

In this paper we show that much higher values can be obtained from
molecules exhibiting Fano resonances. We investigate the
thermoelectric properties of two different molecules, the
1,4-biphenyldithiol (BPDT) and CSW-470-bipyridine (CSW-479)
\cite{wang2006}, as a function of the rotation angles of some of
the molecular groups. Both exhibit Breit-Wigner-like resonances in
the transmission coefficient $\mathcal{T}(E)$ near the Fermi level
corresponding to the HOMO and LUMO levels. However, CSW-479 also
possesses states associated with the bi-pyridine side group, which
are weakly coupled to the backbone and produce an additional Fano
resonance in the electron transmission coefficients
\cite{Car06,papadopoulos2006}. The position of the Fano resonance
can be tuned via a variety of mechanisms, including ring rotation
of the side group \cite{papadopoulos2006} and gating by nearby
polar molecules such as water \cite{water}, whereas Breit-Wigner
resonances are relatively insensitive to such effects. In what
follows, we demonstrate that changing the conformation of the side
group in CSW-479 moves the Fano resonance close to the Fermi
energy and creates huge changes in the magnitudes of the
thermopower and can even change its sign, whereas only small
changes occur when the conformation the BPDT molecule is changed.

\begin{figure}
\includegraphics[width=\columnwidth]{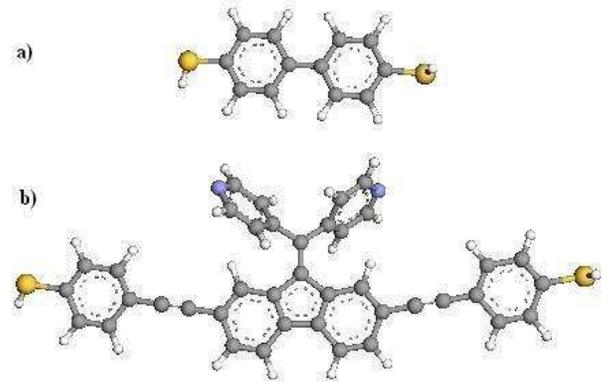}
\caption{\label{fig-molecules}Composition of the studied molecules
a) 1,4-Biphenyl-dithiol (BPDT) and b) CSW-479-bipyridine
(CSW-479)}
\end{figure}

The thermal and electrical properties of a molecular device can be
understood by studying the thermoelectric coefficients $G$, $L$,
$M$ and $K$, which relate the charge current $I$ and heat current
$\dot{Q}$ to the electrical bias $\Delta V$ and the temperature
difference $\Delta T$ across the system. These equations,
$I=G\Delta V+L\Delta T$ and $\dot{Q}=M\Delta V+K\Delta T$, are
commonly rewritten in terms of measurable thermoelectic
coefficients: the electrical conductance $G$, thermopower $S$,
Peltier coefficient $\Pi$ and thermal conductance $\kappa$,
defined by:

\begin{equation}
\left(
\begin{array}{c}
\Delta V\\ \dot{Q}
\end{array}\right)
=
\left(
\begin{array}{cc}
R &S\\\Pi &\kappa
\end{array}\right)
\left(
\begin{array}{c}
I \\ \Delta T
\end{array}\right)
\end{equation}

Thermoelectric properties can therefore be utilized in either
charge-driven cooling devices or heat-driven current generators.
In conventional devices the maximum efficiency of either heat
transfer or current generation is proportional to the
dimensionless thermoelectric figure of merit
$ZT=\frac{S^2GT}{\kappa}$ defined in terms of the measurable
thermoelectric coefficients, where high efficiency corresponds to
values $ZT>>1$. Some of the highest measured and predicted values
of $S$ and $ZT$ are found in materials that have sharp densities
of states (DOS) \cite{mahan1996,hicks1993} such as bismuth
nanowires \cite{cronin2002} or heavily doped semiconductors
\cite{lin2000}. It is therefore expected that similar high values
could be obtained in molecular devices since molecules also have
sharp DOS close to their molecular levels.

The thermoelectric coefficients $S$, $G$ and $\kappa$ can be
computed by extending the Landauer-B\"{u}ttiker formalism to
include both charge and heat currents. In the linear temperature
and bias regime the three required thermoelectric coefficients
\cite{Electric} can be expressed in terms of the moments $L_n$ of
the transmission coefficient  via the following equations
\cite{claughton1996}.

\begin{eqnarray}
S=&&-\frac{1}{e T}\frac{L_1}{L_0}\label{eq-S}\\
G=&&\frac{2e^2}{h}L_0\label{eq-G}\\
\kappa=&&\frac{2}{h}\frac{1}{T}\bigg(L_2-\frac{L_1^2}{L_0}
\bigg)\label{eq-kappa}\\
ZT=&&\frac{1}{\frac{L_0L_2}{L_1^2}-1}\\
\qquad L_n=&&\int_{-\infty}^{\infty} (E-E_f)^n \mathcal{T}(E)
\frac{\partial f(E)} {\partial E}\mathrm{d}E \label{eq-L}
\end{eqnarray}

\noindent If $T(E)$ is a slowly-varying function of $E$, then at
low temperatures, theses expressions can be simplified by
expanding $T(E)$ about $E=E_F$. In the present case, where $T(E)$
varies rapidly when a Fano resonance is present, we calculate the
full integrals rather than taking the low temperature limit.

To calculate thermoelectric properties from first principles, we
use the density functional theory code SIESTA \cite{soler2002}
which employs norm-conserving pseudopotentials to remove the core
electrons \cite{troullier1991} and a local atomic orbital basis
set. In particular we use a double-zeta polarized basis set for
all atoms and the local density approximation for the exchange and
correlation functional as parameterized by Ceperley and Alder
\cite{cep1980}. The Hamiltonian and overlap matrices are
calculated on a real space grid defined by a plane wave cut-off of
200 Ry. Each molecule is relaxed into the optimum geometry until
the forces on the atoms are smaller than 0.02 eV/{\AA} and then
sections of the molecules are rotated to the desired angles. The
contact atoms are attached to the hollow site of a gold (111)
surface consisting of 6 layers with 25 atoms per layer. The
transmission coefficients are found by using the non-equilibrium
Green's function formalism\cite {smeagol}.

The large values of $S$ and $ZT$ predicted below rely only on the
ability to move the Fano resonance close to the Fermi energy and
are independent of the physical mechanism used to produce this
shift. Furthermore, even though DFT may not predict the exact
position of the Breit-Wigner and Fano resonances, the fact that
the side groups are also aromatic groups and have energy levels
close to the energy levels of the backbone, ensures the presence
of Fano resonances near the Fermi level. Therefore an adjustment
to the environment of the side group will generate the required
shift. In what follows we examine the effect of rotating either
the phenyl rings in BPDT or the side group in CSW-479, which could
be induced by tilting the molecule using an STM tip \cite
{Haiss2004} or via steric hindrance
\cite{Ven06,finch2008,venkataraman2006,woitellier1989}.

The sections of the molecules to rotate are chosen so that in BPDT
the coupling along the backbone and in CSW-479 the coupling to the
side group (see Fig. (\ref{fig-molecules})) are changed. For BPDT
the molecule is first relaxed and then the torsion angle between
the phenyl rings is artificially increased from 0$^{\text{o}}$ to
90$^{\text{o}}$, where 0$^{\text{o}}$ corresponds to the case
where the rings are planar. For CSW-479 the side group is rotated
from 0$^{\text{o}}$ to 180$^{\text{o}}$ about the C-C bond that
couples it to the backbone, where 0$^{\text{o}}$ corresponds to
when the bi-pyridine side group is approximately in line with the
molecular backbone.

\begin{figure}
\includegraphics[width=\columnwidth]{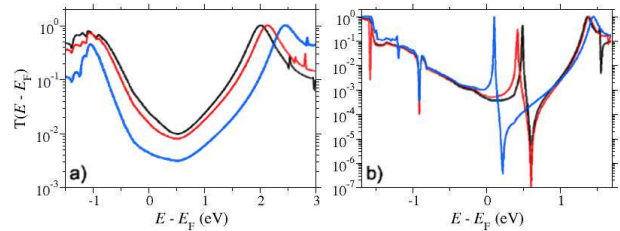}
\caption{\label{fig-transmission}Transmission coefficient for the
a) BPDT and b) CSW-479 molecules for twist angles of
0$^{\text{o}}$ (black), 30$^{\text{o}}$ (red), and 60$^{\text{o}}$
(blue).}
\end{figure}

Fig. (\ref{fig-transmission}) shows  the transmission coefficients
for a) BPDT and b) CSW-479 as a function of energy for different
twist angles when the molecules are connected to the hollow site
on the gold surfaces. As the phenyl rings are rotated from
0$^{\text{o}}$ to 60$^{\text{o}}$ in BPDT, the conjugation of the
molecular backbone is reduced, which moves the LUMO level up in
energy and reduces the transmission at the Fermi energy, following
a known $\mathrm{cos}^2(\theta)$ law
\cite{finch2008,venkataraman2006,woitellier1989}. However the
overall shape of the transmission remains almost unchanged. In
contrast the CSW-479 molecule possesses a Fano resonance near the
Fermi energy and as the side group is rotated, the Fano resonance
moves to a lower energy, leading to a significant change in the
slope of $T(E)$ near $E=E_F$.

\begin{figure}
\includegraphics[width=\columnwidth]{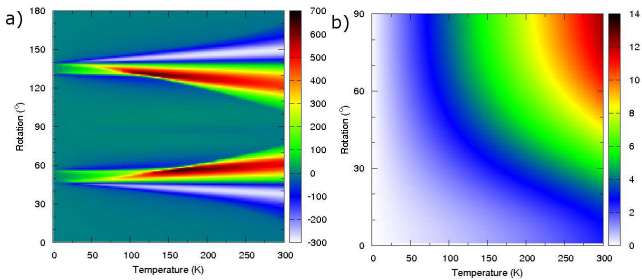}
\caption{\label{fig-thermopower}Contour plot of the thermopower
$S$ for a) CSW-479 and b)BPDT as a function of temperature and
twist angle. Note that the color scales used in the two figures
differ by orders of magnitude.}
\end{figure}

\begin{figure}
\includegraphics[width=\columnwidth]{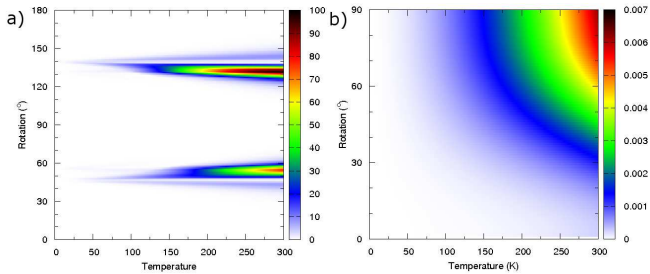}
\caption{\label{fig-figure_of_merit}Contour plot of the figure of
merit $ZT$ for a) CSW-479  and b) BPDT as a function of
temperature and twist angle.}
\end{figure}

For each twist angle,  Fig. (\ref{fig-thermopower}) shows the
thermopower for a) CSW-479 and b) BPDT, obtained by using the
transmission coefficients shown in Fig. (2), to evaluate equations
(\ref{eq-S}) and (\ref{eq-L}). The thermopower in the BPDT system
increases smoothly with rotation angle at high temperatures and
remains positive. This is expected since around the Fermi energy
the transmission coefficient decays slowly for all angles and
therefore the asymmetric term L$_{\text{1}}$ in Eq. (2) is always
negative. The thermopower for the optimized geometry with
$\theta\simeq 35^{\text{o}}$ is calculated to be 6.49
$\mu$VK$^{-1}$, which is close to the value measured in Ref.
\onlinecite{reddy2007}. The discrepancy could be attributed to
differences in the contact geometry. In fact, we know from
previous calculations that by altering the contact configuration
or by varying the distance between the leads and the molecule
\cite{Hof08,Gar07}, the transmission coefficients can change
dramatically, which modifies the slope near the Fermi level.
Predicted thermopowers can also differ from experimental values
even when the contact configuration is correct, due to
approximations built into DFT.

In contrast to BPDT, the CSW-479 molecule produces large
magnitudes for the thermopower when the Fano resonance is within
$k_\mathrm{B}T$ of the Fermi energy. At low temperatures
($T<50K$), the largest value of $S$ is found to be approximately
200 $\mu$VK$^{\text{-1}}$ and at higher temperatures of
approximately 670 $\mu$VK$^{\text{-1}}$. Furthermore the
thermopower becomes negative as the angle increases and the Fano
resonance approaches the Fermi energy from above. This suggests
that thermoelectric properties can be tuned by modifying side
groups and their coupling to the backbone. In this case it is also
possible obtain a negative thermopower at high temperatures even
though the Fermi energy is still inside the HOMO-LUMO gap. The
sign change in the thermopower arises from a combination of the
asymmetric term L$_{\text{1}}$ and the intrinsic asymmetric shape
of the Fano resonance. Therefore, the idea that the hole or
electron-like character of conduction is determined by the sign of
the thermopower is ambiguous for systems with Fano resonances
close to the Fermi energy.

As a demonstration of the generic nature of these predictions, we
repeated the above calculations for molecules connected to a gold
atom on top of a gold electrode. The results were qualitatively
similar, even though the details changed, such as the position of
the Fermi energy, which shifted slightly closer to the HOMO level
due to the reduction of the charge transfer in the top
configuration.

Fig. (\ref{fig-figure_of_merit}) shows the calculated
dimensionless figures of merit for a) CSW-479 and b) BPDT. Due to
the low thermopower and conductance near the Fermi level the
figure of merit for BPDT is low for all rotation angles, and
therefore this molecule has a low efficiency of thermoelectric
heat transfer. This is not the case for CSW-479, where even at low
temperatures a figure of merit $ZT>>1$ can be achieved for
rotation angles where the Fano resonance is close to the Fermi
energy. This demonstrates that the temperature difference can be
mechanically tuned by adjusting the coupling between the localized
states and the backbone. Such large figures of merit  have been
predicted for semi-conducting carbon nano-tubes
\cite{esfarjani2006}. However to date this is the largest
predicted figure of merit value for an un-gated single molecular
device.

We have shown there is a huge difference in the thermoelectric
properties of molecules with and without side groups. In contrast
with electrical properties, the thermoelectric properties of a
molecular device depend on the shape of the transmission profile
about the Fermi energy rather than its absolute magnitude. By
moving a Fano resonance through the Fermi energy, the shape of the
transmission can be dramatically altered to produce both huge
thermopower and figures of merit, corresponding to a
high-efficiency thermoelectric cooling device. In practice, at
high temperatures, this value could be limited by the contribution
to the thermal conductance from phonons \cite{Mur08}, which acts
in parallel to the electronic contribution to the thermal
conductance and is not included here. Electron transport through
quasi-one-dimensional wires is also sensitive to disorder
\cite{l1,l2}, which could  arise under ambient conditions, due to
the presence of polar water molecules \cite{water}.

\begin{acknowledgments}
We would like acknowledge funding from the EPSRC, the NWDA, the
DTI and the MCRTN 'FUNMOLS'.
\end{acknowledgments}


\begin{thebibliography}{99}
\bibitem{weber2002}
H. B. Weber, J. Reichert, F. Weigend, R. Ochs, D. Beckmann, M.
Mayor, R. Ahlrichs and H. v. L\"{o}hneysen, Chem. Phys. {\bf 281},
113 (2002).

\bibitem{xu2003}
B. Xu and N. J. Tao, Science {\bf 301}, 1221 (2003).

\bibitem{Haiss2004}
Wolfgang Haiss, Changsheng Wang, Iain Grace, Andrei S. Batsanov,
David J. Schiffrin, Simon J. Higgins, Martin R. Bryce, Colin J.
Lambert and Richard J. Nichols, Nature Materials, {\bf 5} 995
(2006).

\bibitem{choi2006}
B.-Y. Choi, S-J. Kahng, S. Kim, H. Kim,  H. W. Kim, Y. J. Song, J.
Ihm and Y. Kuk, Phys. Rev. Lett. {\bf 96}, 156106 (2006).

\bibitem{lortscher2006}
E. L\"{o}rtscher, J. W. Ciszek, J.Tour and H. Reil, Small {\bf 2},
973 (2006).

\bibitem{ashwell2005}
G. J. Ashwell, A. Mohib and J. R. Miller, J. Mater. Chem. {\bf
15}, 1160 (2005).

\bibitem{ashwell2006}
G. J. Ashwell, W. D. Tyrrell, B. Urasinska, C. Wang and M. R.
Bryce, Chem. Commun. {\bf 15}, 1640 (2006).

\bibitem{he2006}
J. He, B. Chen, A. K. Flatt, J. J. Stephenson, C. D. Doyle and J.
M. Tour, Nature Materials {\bf 5}, 63 (2006).

\bibitem{long2006}
D. P. Long, J. L. Lazorcik, B. A. Mantooth, M. H. Moore, M. A.
Ratner, A. Troisi, Y. Yao, J. W. Ciszek, J. M. Tour and R.
Shashidhar, Nat. Mat. {\bf 5}, 901 (2006).

\bibitem{Koh04}
J. Koch, F. von Oppen, Y. Oreg, and E. Sela, Phys. Rev. B {\bf
70}, 195107 (2004).

\bibitem{reddy2007}
P. Reddy, S. Jang, R. A. Segalman and A. Majumdar, Science {\bf
315}, 1568 (2007).

\bibitem{paulsson2003}
M. Paulsson and S. Datta, Phys. Rev. B {\bf 67}, 241403(R) (2003).

\bibitem{pauly2008}
F. Pauly, J. K. Viljas, and J. C. Cuevas, Phys. Rev. B {\bf 78},
035315 (2008).

\bibitem{wang2005}
B. Wang, Y. Xing, L. Wan, Y. Wei and J. Wang, Phys. Rev. B {\bf
71}, 233406 (2005).

\bibitem{zheng2004}
X. Zheng, W. Zheng, Y. Wei, Z. Zeng and J. Wang, J. Chem. Phys.
{\bf 121}, 8537 (2004).

\bibitem{wang2006}
C. Wang, A. S. Batsanov and M. R. Bryce, J. Org. Chem. {\bf 71},
108 (2006).

\bibitem{Car06}
D. M. Cardamone, C. A. Stafford, and S. Mazumdar, Nano Lett. {\bf
6}, 2422 (2006).

\bibitem{papadopoulos2006}
T. A. Papadopoulos, I. M. Grace and C. J. Lambert, Phys. Rev. B
{\bf 74}, 193306 (2006).

\bibitem{water} Leary {\em et al.}, preprint (2008).

\bibitem{mahan1996}
G. D. Mahan and J. O. Sofo, Proc. Natl. Acad. Sci. {\bf 93}, 7436
(1996).

\bibitem{hicks1993}
L. D. Hicks and M. S. Dresselhaus, Phys. Rev. B {\bf 47}, 16631
(1993).

\bibitem{cronin2002}
S. B. Cronin, Y.-M. Lin, O. Rabin and M. S. Desselhaus, 21st
International Conference on Thermoelectrics: Symposium Proc., 243
(2002).

\bibitem{lin2000}
Y.-M. Lin, X. Sun and M. S. Dresselhaus, Phys. Rev. B {\bf 62},
4610 (2000).

\bibitem{Electric}
Here, $\kappa$ includes only the electric contribution to the
thermal conductance, with the phonon term neglected as in Ref.
\onlinecite{wang2005}.

\bibitem{claughton1996}
N. R. Claughton and C. J. Lambert, Phys. Rev. B {\bf 53}, 6605
(1996).

\bibitem{soler2002}
J. M. Soler, E. Artacho, J. D. Gale, A. Garcia, J. Junquera,
P.Ordejon and D. Sanchez-Portal, J. Phys.: Condens Matter. {\bf
14}, 2745 (2002).

\bibitem{troullier1991}
N. Troullier and J. L. Martins, Phys. Rev. B {\bf 43}, 1993
(1991).

\bibitem{cep1980}
D. M. Ceperley and B. J. Alder, Phys. Rev. Lett. {\bf 45}, 566
(1980).

\bibitem{smeagol}
A. R. Rocha, V. M. Garc\'{\i}a-Su\'arez, S. Bailey, C. Lambert, J.
Ferrer, and S. Sanvito,  Phys. Rev. {\bf B 73}, 085414 (2006).

\bibitem{Ven06}
L. Venkataraman, J. E. Klare, C. Nuckolls, M. S. Hybersten, and M.
L. Steigerwald, Nature (London) {\bf 442}, 904 (2006).

\bibitem{finch2008}
C. M. Finch, S. Sirichantaropass, S. W. Bailey, I. M. Grace, V. M.
Garc\'{\i}a-Su\'arez and C. J. Lambert, J. Phys.: Condens. Matter
{\bf 20}, 022203 (2008).

\bibitem{venkataraman2006}
L. Venkataraman, J. E. Klare, C. Nuckolls, M. S. Hybertsen and M.
L. Steigerwald, Nature (London) {\bf 442}, 904 (2006).

\bibitem{woitellier1989}
S. Woitellier, J. P. Launay and C. Joachim, Chem. Phys. {\bf 131},
481 (1989).

\bibitem{Gar07}
V. M. Garc\'{\i}a-Su\'arez, T. Kostyrko, S. Bailey, C. Lambert,
and B. R. Bu\l ka, Phys. Stat. Sol. (b) {\bf 244}, 2443 (2007).

\bibitem{Hof08}
R. C. Hoft, M. J. Ford, V. M. Garc\'{\i}a-Su\'arez, C. J. Lambert
and M. B. Cortie, J. Phys.: Condens. Matter {\bf 20}, 025207
(2007).

\bibitem{Mur08}
P. Murphy, S. Mukerjee, and J. Moore, arXiv:0805.3374 (2008).

\bibitem{esfarjani2006}
K. Esfarjani, M. Zebarjadi and Y. Kawazoe, Phys. Rev. B {\bf 73},
085406 (2006).

\bibitem{l1}
C. J. Lambert, J. Phys. C17, 2401 (1984).

\bibitem{l2}
C. J. Lambert, Phys. Rev. B29, 1091 (1984).

\end{thebibliography}
\end{document}